\documentclass[prl,twocolumn]{revtex4}

\usepackage[english]{babel}
\usepackage{amsmath}
\usepackage{epsfig}

\newcommand{\inputEps}[3]{
            \centerline{\epsfxsize=#1 \epsfbox{#3} }
\vskip 12pt {\center\small{#2}} \vskip 18pt}

\begin{document}
\title{\bf Dynamical suppression of radiative decay via atomic deflection
by a standing light wave}

\author{M.A.~Efremov}
%\email{efremov@gon.ran.gpi.ru}
\author{M.V.~Fedorov}
%\email{fedorov@gon.ran.gpi.ru}
\affiliation{ General Physics
Institute, RAS, 119991 Moscow, Russia}
\author{V.P.~Yakovlev}
%\email{yakovlev@theor.mephi.ru}
\affiliation{Moscow State Engineering Physics Institute, 115409
Moscow,Russia}
\author{W.P.~Schleich}
%\email{Wolfgang.Schleich@physik.uni-ulm.de}
\affiliation{ Abteilung f$\ddot{u}$r Quantenphysik,
Universit$\ddot{a}$t Ulm, 89069 Ulm, Germany}

\date{\today}

\begin{abstract}
We consider the radiative decay of atoms scattered by a resonant
standing light wave. Scattering is shown to suppress the Rabi
oscillations and to slow down the atomic radiative decay giving
rise to a power law behavior of the time-dependent level
populations rather than the exponential one.
\end{abstract}

\pacs{ }

\maketitle

\section{Introduction}
Scattering of atoms by a resonant standing light wave is one of
the basic phenomena of the atom optics. The number of publications
on this subject is huge (see, e.g., the books \cite{Ya-Su,
Balykin, Sch} and the references therein). A very special place in
this field belongs to a series of works on atoms with a wide
excited resonance level, the width of which is determined mainly
by spontaneous radiative transitions to nonresonant atomic levels
\cite{ChY, NZ, Z, Batelaan, BO'D, Z2}. Realizability of such a
scheme was demonstrated in the experiment \cite{Z, Z2}. Ar* atoms
were prepared initially in a metastable state $|m\rangle$ ($|1s_5,
J=2\rangle\equiv 4s[3/2]_2$ state, correspondingly, in the Paschen
and standard spectroscopic notations \cite{Z2}). In the
experiment, the resonant light field of the wavelength 801 nm
coupled the level $E_m$ with the excited level $E_e$, which
corresponded to the state $|2p_8, J=2\rangle\equiv 4p[5/2]_2$.
70\% of its width was determined by transitions to levels
different from $E_m$ and, finally, to the ground level $E_g$ (Fig.
1). In principle, in Ar* there is another transition, $|1s_5,
J=2\rangle\rightarrow|2p_4, J=2\rangle$, in which 98\% of the
width of the excited level is determined by spontaneous decay to
levels different from $E_m$ \cite{Batelaan}. In any case, for both
transitions the model of a two-level system with a wide excited
level decaying predominantly to levels different from the
metastable one works reasonably well. This is the model to be
considered in this work.

Both in the experiment \cite{Z, Z2} and, most often, in the theory
\cite{Batelaan}, the investigated regimes of scattering
corresponded to the weak-scattering Bragg regime. This means that
the atomic-beam incidence angle was close or equal to the Bragg
angle and the resonance coupling  was not too strong. In terms of
the Rabi frequency $\Omega$ and the width of the excited level
$\Gamma$, the last assumption implies that $|\Omega|\ll\Gamma$. In
this case, the decay dynamics of an atomic system was shown to
obey the usual exponential law \cite{Z2, Batelaan}. Under some
special conditions the effects like population trapping were
predicted to take place \cite{Batelaan}. But, again, the residual
atomic population was shown to approach its asymptotic non-zero
level exponentially \cite{Batelaan}.

In this work we will investigate the dynamics of spontaneous decay
in the system under consideration at different conditions. First,
we will consider the case of strong Rabi coupling,
$|\Omega|\gg\Gamma$ and $|\Omega|t\gg1$. And, second, we will
consider the case of normal (or almost normal) incidence of atoms
upon the standing wave. This is the diffraction regime of
scattering, in which many diffraction maxima can arise from the
initially well collimated atomic beam.

By investigating the decay dynamics in such a regime we find that
the total time-dependent populations of both metastable and
excited levels fall non-exponentially. In contrast to standard
predictions, the atomic populations are characterized  by
power-law dependencies on the interaction time $t$. The effect is
not connected with formation of any kind of grey or dark states
(as in Ref. \cite{Batelaan}) because asymptotically, at very long
time, atomic populations tend to zero. But, owing to scattering,
the radiative decay appears to be slowed down. In addition, we
find that in atoms scattered by a standing light wave the Rabi
oscillations of atomic populations appear to be strongly
suppressed compared to a pure two-level system in a resonance
field. The physical interpretation of these effects is given.

\section{General equations}
The total wave function $\Psi$ of an atom interacting with a light
field depends on the atomic center-of-mass position vector ${\bf
r}$, intra-atomic variables, and time $t$. The wave function obeys
the Schr$\ddot{\rm o}$dinger equation
\begin{equation}
 \label{shrodinger}
 i\frac{\partial\Psi}{\partial
 t}=\left\{-\frac{1}{2m}\nabla^2+H_{at}-{\bf d}\cdot{\bf E}({\bf r},
 t)\right\}\Psi,
\end{equation}
where $\hbar=1$, $\nabla=\partial/\partial {\bf r}$, ${\bf d}$ is
the intra-atomic dipole moment, and ${\bf E}({\bf r}, t)$ is the
electric field strength, for a standing light wave given by
\begin{equation}
 \label{light-field}
 {\bf E}({\bf r}, t)=2{\bf E}_{0}\cos(\omega t)\cos(kx).
\end{equation}
Here and below $x$ is the center-of-mass coordinate along the axis
parallel to ${\bf k}$; ${\bf k}$ and ${\bf E}_0$ are the wave
vector and field-strength amplitude of one of the two identical
counter-propagating travelling waves forming a standing light wave
(Fig. 1), $k=\omega/c$.

\vskip 12pt

\inputEps{220pt}{Figure 1. A scheme of atom-light scattering and the
internal structure of atomic levels.}{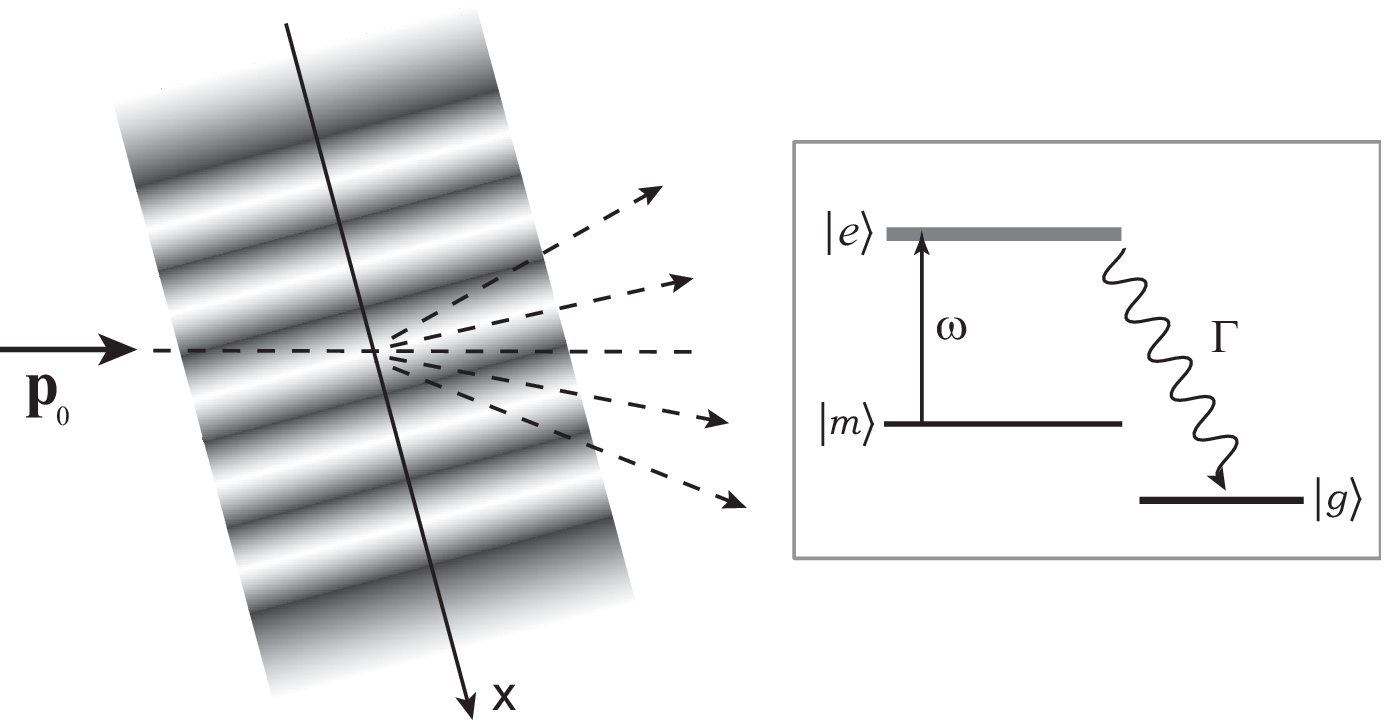}

With respect to intra-atomic variable, the wave function $\Psi$
can be expanded in a series of intra-atomic field-free wave
functions $|i\rangle$
\begin{equation}
 \label{expansion}
 \Psi=\sum_i \varphi_i(x,t )|i\rangle
 \exp\left(-i\,E_{i}t+i{\bf p}_0\cdot{\bf r} -i\frac{p_0^2
 t}{2m}\right),
\end{equation}
where ${\bf p}_0$ is the unperturbed-atom center-of-mass momentum.

In this work we will consider  only the resonance case, when the
light frequency equals the energy spacing between some two
discrete nondegenerate atomic levels, $E_e$ and $E_m$,
$\omega=E_e-E_m$. We assume that $E_m$ and $E_e$ are,
respectively, an infinitely narrow metastable and a wide excited
atomic levels, and the width $\Gamma$ of the excited level is
determined predominantly by its spontaneous decay to the ground
atomic level (inset of Fig. 1).

In the resonance case we keep only two terms in the expansion
(\ref{expansion}) with $i=m$ and $i=e$. Moreover, in the rotating
wave approximation we retain only one of the two terms in the
Euler expansion for cosine $\cos{(\omega
t)}=\frac{1}{2}[\exp{(i\omega t)}+\exp{(-i\omega t)}]$ to drop the
fast oscillating terms $\propto \exp{(\pm i\omega t)}$. The
arising equations for the metastable- and excited-state
center-of-mass wave functions $\varphi_m(x, t)$ and $\varphi_e(x,
t)$ can be written in the form of a matrix Schr${\rm
\ddot{o}}$dinger-like equation for the two-component function
\begin{equation}
 \label{2-comp}
 \Phi(x, t)
 =\left\{\varphi_{m}(x,t)\atop{\varphi_{e}(x,t)}\right\}:
\end{equation}

\begin{equation}
 \label{eq for Phi}
 i\frac{\partial\Phi(x, t)}{\partial t}={\bf H}\,\Phi(x, t)
\end{equation}
with the matrix Hamiltonian
\begin{equation}
 \label{Ham}
  \bf H=\begin{pmatrix}
         -\frac{1}{2m}\nabla_x^2-\frac{i}{m}p_{0x}\nabla_x   & -\frac{1}{2}\Omega\cos(kx) \\
         -\frac{1}{2}\Omega^*\cos(kx) & -\frac{1}{2m}\nabla_x^2-\frac{i}{m}p_{0x}\nabla_x
          -\frac{i\Gamma}{2}
         \end{pmatrix}.
\end{equation}
Here and below $\nabla_x\equiv\partial/\partial x$, $\Omega=2{\bf
d}_{me}\cdot{\bf E}_{0}$ is the Rabi frequency, and ${\bf
d}_{me}\equiv \langle m|{\bf d}|e \rangle$ is the dipole matrix
element.

\section{Adiabatic approximation}

Explicitly, equations for $\varphi_{m}(x,t)$ and
$\varphi_{e}(x,t)$ equivalent to (\ref{eq for Phi}) have the form
(in the case of normal incidence, $p_{0x}=0$):
\begin{equation}
 \label{eq-m-gen}
 i\frac{\partial}{\partial t}\varphi_m(x, t)=-\frac{1}{2m}\nabla_x^2\varphi_m(x, t)
 -\frac{\Omega}{2}\cos(kx)\,\varphi_e(x, t)
\end{equation}
and
$$
 i\frac{\partial}{\partial t}\varphi_e(x, t)=
 -\frac{1}{2m}\nabla_x^2\varphi_e(x, t)
$$
\begin{equation}
 \label{eq-e-gen}
  -i\frac{\Gamma}{2}\varphi_e(x, t)
 -\frac{\Omega^{*}}{2}\cos(kx)\,\varphi_m(x, t).
\end{equation}
In this work we will use the adiabatic approximation in which we
will drop the kinetic energy operator $-\nabla_x^2/2m$ in the
Hamiltonian ${\bf H}$ (\ref{Ham}) and, hence, the terms
proportional to $-\nabla_x^2/2m$ on the right-hand side of Eqs.
(\ref{eq-m-gen}) and (\ref{eq-e-gen}) to get
\begin{equation}
 \label{matrix}
 {\bf H}\approx{\bf H}_{ad}=
 \begin{pmatrix} 0 & -\frac{1}{2}\Omega\cos(kx) \\
  -\frac{1}{2}\Omega^{*}\cos(kx) & -\frac{i}{2}\Gamma
 \end{pmatrix},
\end{equation}
\begin{equation}
 \label{eq-for-phi-m}
 i\frac{\partial}{\partial t}\varphi_m(x, t)=
 -\frac{\Omega}{2}\cos(kx)\,\varphi_e(x, t),
\end{equation}
and
\begin{equation}
 \label{eq-for-phi-e}
 i\frac{\partial}{\partial t}\varphi_e(x, t)= -i\frac{\Gamma}{2}\varphi_e(x, t)
 -\frac{\Omega^{*}}{2}\cos(kx)\,\varphi_m(x, t).
\end{equation}
For the kinetic energy term $(-\nabla_x^2/2m)\varphi_m(x, t)$ to
be dropped from Eq. (\ref{eq-m-gen}), it must be smaller than the
other two terms retained in Eq. (\ref{eq-for-phi-m}). Hence,
qualitatively, the applicability criterion for Eq.
(\ref{eq-for-phi-m}) has the form
\begin{equation}
 \label{criterion}
 \left\langle\frac{-\nabla_x^2}{2m}\right\rangle\ll
 \left\langle\frac{\partial}{\partial  t}\right\rangle,
\end{equation}
where angular brackets denote averaging over the state $\Psi$. As
for Eqs. (\ref{eq-e-gen}) and (\ref{eq-for-phi-e}), they contain
additional terms $\propto\Gamma$, which can be large, and the
adiabaticity criterion for Eq. (\ref{eq-for-phi-e}) is given by
\begin{equation}
 \label{criterion2}
 \left\langle\frac{-\nabla_x^2}{2m}\right\rangle\ll
 \max\left\{\Gamma,\;\left\langle\frac{\partial}{\partial
 t}\right\rangle\right\}.
\end{equation}
By comparing the conditions of Eqs. (\ref{criterion}) and
(\ref{criterion2}), we see that the first of them is sufficient
for the kinetic energy operator to be dropped from the Hamiltonian
${\bf H}$ (\ref{Ham}) and for the adiabatic approximation to be
valid. An explicit form of the criterion (\ref{criterion}) is
discussed below.

If $\Gamma\gg\left\langle\partial/\partial t\right\rangle$, the
condition (\ref{criterion}) can be invalid whereas, still, the
condition (\ref{criterion2}) can be satisfied. In this case the
kinetic energy operator in Eq. (\ref{eq-m-gen}) has to be retained
whereas in Eq. (\ref{eq-e-gen}) it can be dropped, and the
equations to be solved are (\ref{eq-m-gen}) and
(\ref{eq-for-phi-e}). Such a generalization of the adiabatic
approximation will be considered elsewhere.

By expressing $\varphi_e(x, t)$ from Eq. (\ref{eq-for-phi-m}) via
$\varphi_m(x, t)$, $\varphi_e(x,
t)=-2i\dot{\varphi}_m/\Omega\cos{(kx})$, and substituting
$\varphi_e(x, t)$ into Eq. (\ref{eq-for-phi-e}), we can reduce the
two equations (\ref{eq-for-phi-m}) and (\ref{eq-for-phi-e}) to a
single second-order equation for $\varphi_m(x, t)$
\begin{equation}
 \label{second-order}
 \ddot{\varphi}_m + \frac{\Gamma}{2}\dot{\varphi}_m +
 \frac{|\Omega|^2\cos^2{(kx)}}{4} \varphi_m=0.
\end{equation}

\section{Quasienergy solutions}
As the Hamiltonian (\ref{Ham}) is stationary, Eq. ({\ref{eq for
Phi}}) has solutions of the form
\begin{equation}
 \label{qe}
 \Phi(x, t) =\exp(-i\gamma t)\,u_{\gamma}(x),
\end{equation}
where $\gamma$ and $u_{\gamma}$ are the quasienergies and
quasienergy wave functions to be found from the eigenvalue
equation
\begin{equation}
\label{matrix-equation-qasienergy}
 {\bf H}_{ad}\,u_{\gamma}(x) =\gamma\,u_{\gamma}(x).
\end{equation}
In the adiabatic approximation solution of the quasienergy problem
is very simple. As the operator ${\bf H}_{ad}$ (\ref{matrix}) does
not contain any derivatives over $x$, the coordinate $x$ plays the
role of a parameter or a quantum number, and the
position-dependent eigenvalues of ${\bf H}_{ad}$ are easily found
to be given by
\begin{equation}
\label{quasienergies} \gamma_\pm
(x)=-i\frac{\Gamma}{4}\pm\frac{1}{2}\sqrt{-\frac{\Gamma^2}{4}
+|\Omega|^2\cos^2(kx)}.
\end{equation}

\noindent The quasienergy $\gamma_{+}(x)$ is a direct
generalization of the Chudesnikov-Yakovlev complex potential
\cite{ChY} (at zero detuning), which follows from Eq.
(\ref{quasienergies}) in the limit $|\Omega|\ll \Gamma$
\begin{equation}
 \label{Ch-Y}
 \gamma_+(x)\approx -i\frac{|\Omega|^2 \cos^2(kx)}{2\Gamma}
 \equiv V_{CY}(x)|_{\omega=E_e-E_m}.
\end{equation}
In a general case, the complex quasienergies $\gamma_\pm(x)$
(\ref{quasienergies}) determine both average 'center-of-mass'
values ${\rm Re}(\gamma_\pm)$ and widths $\Gamma_\pm = -2{\rm
Im}(\gamma_\pm)$ of the atomic quasienergy levels. Explicitly, the
widths $\Gamma_\pm (x)$  are given by
\begin{equation}
 \label{widths}
 \Gamma_\pm (x)=\frac{\Gamma}{2}\mp
 {\rm Re}\left({\sqrt{\frac{\Gamma^2}{4}-|\Omega|^2\cos^2(kx)}}\right)
\end{equation}
The broadened quasienergy levels or zones are described in Fig. 2.
The curves at these picture correspond to

\vskip 12pt

\inputEps{180pt}{Figure 2. Quasienergy zones of an atom in
a standing light wave  in the cases $2|\Omega|/\Gamma=1/\sqrt{2}$
(a) and $2\sqrt{2}$ (b).}{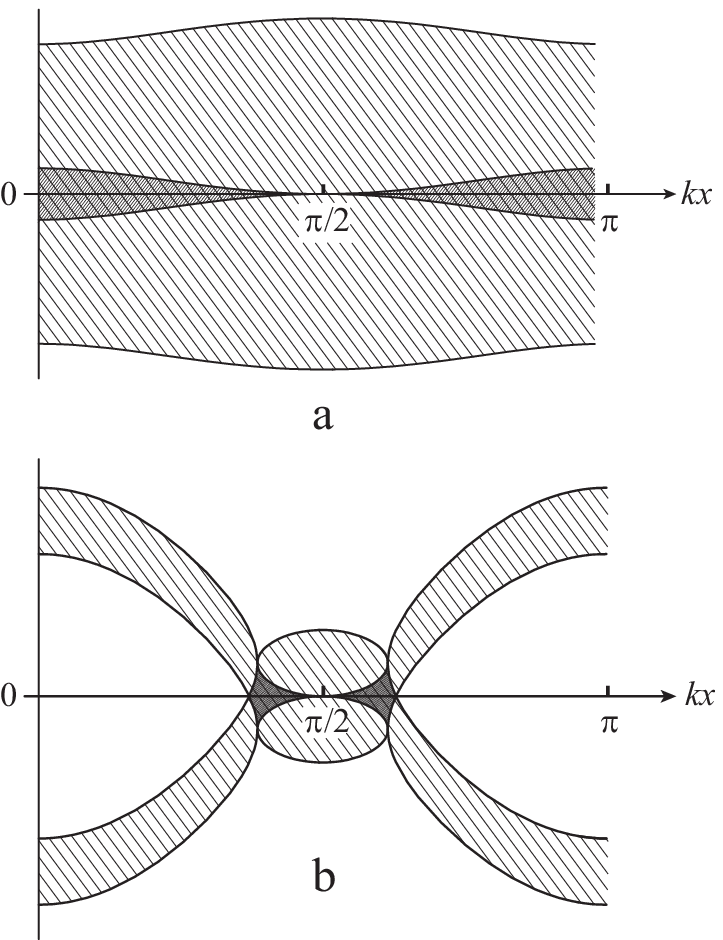}

\noindent boundaries of zones determined as ${\rm Re}[\gamma_\pm
(x)] +\frac{1}{2}\Gamma_\pm (x)$ and ${\rm Re} [\gamma_\pm (x)]
-\frac{1}{2}\Gamma_\pm (x)$. The spacings between the boundaries
are equal to the widths of the zones $\Gamma_+(x)$. Two different
zones are indicated by different shading. The pictures (a) and (b)
correspond to weak ($2|\Omega|<\Gamma$) and strong
($2|\Omega|>\Gamma$) resonance or Rabi coupling of levels $E_m$
and $E_e$. In the case of weak Rabi coupling, one of the zones is
much narrower than another, and the wide zone can be eliminated
adiabatically to give rise to the description in terms of the
potential $V_{CY}(x)$ (\ref{Ch-Y}). Mathematically such an
elimination of a wide quasienergy zone is equivalent to dropping
the second-order derivative term in Eq. (\ref{second-order}),
which gives
\begin{equation}
 \label{Chu-Ya}
 i\frac{\partial}{\partial t}\varphi_m(x, t)=V_{CY}(x)\varphi_m(x,
 t).
\end{equation}
The condition under which the second-order derivative in Eq.
(\ref{second-order}) can be dropped is easily estimated with the
help of Eq. (\ref{Chu-Ya}): $\ddot{\varphi}_m\sim
V_{CY}\dot{\varphi}_m\sim V_{CY}^2{\varphi}_m$. Hence,
$\ddot{\varphi}_m\ll\Gamma \dot{\varphi}_m$ if $|V_{CY}|\ll\Gamma$
or $|\Omega\cos{(kx)|\ll\Gamma}$.

In the case of strong Rabi coupling such adiabatic approximation
and all the resulting equations can be invalid at $x$ close to the
branching points of the root square in Eq. (\ref{quasienergies}).
Indeed, at these points the derivative $\nabla_x$ becomes
infinitely large and the kinetic energy terms in Eqs. (\ref{Ham}),
(\ref{eq-m-gen}), (\ref{eq-e-gen} ) cannot be dropped. But in the
region of $x$ close to $\pi/2k$ (which is most important for the
given below long-time analysis) even in the case
$|\Omega|\gg\Gamma$ the width $\Gamma_{+}(x)$ (\ref{widths}) is
very small and can be approximated by a parabolic dependence on
$x-\pi/2k$
\begin{equation}
 \label{gamma+}
 \Gamma_{+}\approx -2{\rm Im}\left(V_{CY}\right)
 =\frac{|\Omega|^2 \cos^2(kx)}{\Gamma}
 \approx\frac{|\Omega|^2 k^2}{\Gamma}\left(x-\frac{\pi}{2k}\right)^2.
\end{equation}
As in this region $\Gamma_{-}\approx\Gamma\gg\Gamma_{+}$, we get
again a narrow quasienergy zone at the background of a wide one,
and again adiabatic elimination appears to be applicable.

The role of "nonadiabatic" points, where $\Gamma \approx 2|\Omega
\cos{{kx}}|$, will be discussed elsewhere. In principle, the
arising peculiarities can be observed in experiments with
scattering of narrow atomic wave packets aimed specifically at
these points. Such a formulation of the problem will be discussed
separately too.

\section{Solution of the initial-value problem}
The found above quasienergies  $\gamma_\pm (x)$
(\ref{quasienergies}) are sufficient for solving the initial-value
problem. By assuming that the interaction is turned on suddenly at
$t=0$ and that $\varphi_m(x,0)=1$ and $\varphi_e(x,0)=0$, we
present the time-dependent functions $\varphi_{m,e}(x, t)$ in the
form of superpositions
\begin{equation}
 \label{in-value}
 \varphi_{m,e}(x,t)=\sum_\pm A_{m,e}^{(\pm)}(x)\exp\left({-i\gamma_\pm (x)t}\right),
\end{equation}
where the coefficients $A_{m,e}^{(\pm)}(x)$ are to be found from
the initial conditions, which yield
\begin{equation}
 \label{in cond}
 A_m^{(+)}+A_m^{(-)}=1,\;\;\;A_e^{(+)}+A_e^{(-)}=0,
\end{equation}
and the equations following from  Eq. (\ref{eq for Phi}) (with the
Hamiltonian (\ref{matrix}))
\begin{equation}
 \label{e-m connection}
 A_e^{(\pm)}=-\frac{2\gamma_\pm}{\Omega\cos{(kx)}}\,A_m^{(\pm)}.
\end{equation}
Eqs. (\ref{in cond}) and (\ref{e-m connection}) are solved easily
to give
\begin{equation}
 \label{A_m}
 A_m^{(\pm)}=\mp\frac{2\gamma_{\mp}}{\sqrt{-\Gamma^2+4|\Omega|^2\cos^2{(kx)}}}
 \end{equation}
and
\begin{equation}
 \label{A_e}
  A_e^{(\pm)}=\mp\frac{\Omega^*\cos{(kx)}}
 {\sqrt{-\Gamma^2+4|\Omega|^2\cos^2{(kx)}}}.
\end{equation}
The corresponding time-dependent center-of-mass atomic wave
functions are given by
\begin{equation}
 \label{fi_m}
 \varphi_m(x,t)=\sum_\pm \mp\frac{2\gamma_{\mp}(x)\exp(-i\gamma_\pm (x) t)}
 {\sqrt{-\Gamma^2+4|\Omega|^2\cos^2(kx)}}
\end{equation}
and
\begin{equation}
 \label{fi_e}
 \varphi_e(x,t)=\sum_\mp \mp\frac{ \Omega^*\cos(kx)\exp(-i\gamma_\pm (x) t)}
 {\sqrt{-\Gamma^2+4|\Omega|^2\cos^2(kx)}}.
\end{equation}
The squared absolute values of the functions $\varphi_{m,e}(x,t)$
determine the probability densities to find an atom at a time $t$
in a vicinity of a point $x$ at the levels $E_m$ and $E_e$
\begin{equation}
 \label{densities}
 \frac{dW_{m,e}(x, t)}{dx}
 =\frac{k}{\pi}\,\left|\varphi_{m,e}(x,t)\right|^2.
\end{equation}
Integrated over $x$ from zero to $\pi/k$, the probability
densities $dW_{m,e}(x, t)/dx$ give the time-dependent total
probabilities of scattering at a single period of a standing light
wave
$$
 W_{tot}^{(m,e)}(t)=
 \int_0^{\pi/k}dx\frac{dW_{m,e}(x, t)}{dx}
$$
\begin{equation}
 \label{total-prob}
 =\frac{k}{\pi}\int_0^{\pi/k}dx \mid \varphi_{m,e}(x,t)\mid^2.
\end{equation}

\section{Long-time asymptotic limit}
The long-time asymptotic limit corresponds to the case $\Gamma
t\gg 1$. In this case the main contribution to the integral over
$x$ in Eq. (\ref{total-prob}) is given by the most slowly decaying
terms. In the case of a strong Rabi coupling $|\Omega|\gg\Gamma$
such slowly decaying terms correspond to the quasienergy
$\gamma_{+}(x)$ and to the region of $x$ close to $\pi/2k$. By
assuming that in this region the product $|\Omega|^2\cos^2{(kx)}$
is small compared to $\Gamma^2/4$, we can reduce Eqs. (\ref{fi_m})
and (\ref{fi_e}) to the form
\begin{equation}
 \label{phi_m_smallW^2/G}
  |\varphi_m(x,t)|^2\approx \exp{\left\{-\frac{|\Omega|^2 t}{\Gamma}\cos^2{(kx)
 }\right\}}
\end{equation}
and
\begin{equation}
 \label{phi_e_smallW^2/G}
 |\varphi_e(x,t)|^2\approx \frac{|\Omega|^2}{\Gamma^2}
 \cos^2{(kx)}\exp{\left\{-\frac{|\Omega|^2 t}{\Gamma}\cos^2{(kx)
 }\right\}}.
\end{equation}
These equations follows also from Eq. (\ref{second-order}), which
is solved easily and which is valid under the condition
$|\Omega\cos{(kx)}|<2\Gamma$. The position-dependent
metastable-state probability density was found earlier (Eq. (19)
of Ref. \cite{NZ} in which the spontaneous decay rate $\gamma$
should be substituted by $|\Omega|^2/2\Gamma$).

At $x$ close to $\pi/2k$ and with $\cos{(kx)}$ approximated by the
parabolic function (\ref{gamma+}), Eqs. (\ref{phi_m_smallW^2/G})
and (\ref{phi_e_smallW^2/G}) take the form
\begin{equation}
 \label{phi_m_Gauss}
  |\varphi_m(x,t)|^2\approx
  \exp{\left\{-\frac{|\Omega|^2 t}{\Gamma}\left(kx-\frac{\pi}{2}\right)^2
  \right\}}
\end{equation}
and
\begin{equation}
 \label{phi_e_Gauss}
 |\varphi_e(x,t)|^2\approx \frac{|\Omega|^2}{\Gamma^2}
 \left(kx-\frac{\pi}{2}\right)^2\exp{\left\{-\frac{|\Omega|^2 t}
 {\Gamma}\left(kx-\frac{\pi}{2}\right)^2
\right\}}.
\end{equation}
The functions $|\varphi_m(x,t)|^2$ (\ref{phi_m_Gauss}) and
$|\varphi_e(x,t)|^2$ (\ref{phi_e_Gauss}) are plotted in Fig. 3.
The width $\Delta x$ of the interval where

\vskip 18pt

\inputEps{200pt}
{Figure 3. Probability densities to find an atom at the levels
$E_m$ and $E_e$ around a  position $x$; $|\Omega|/\Gamma=3$,
$\Gamma t=2$.}{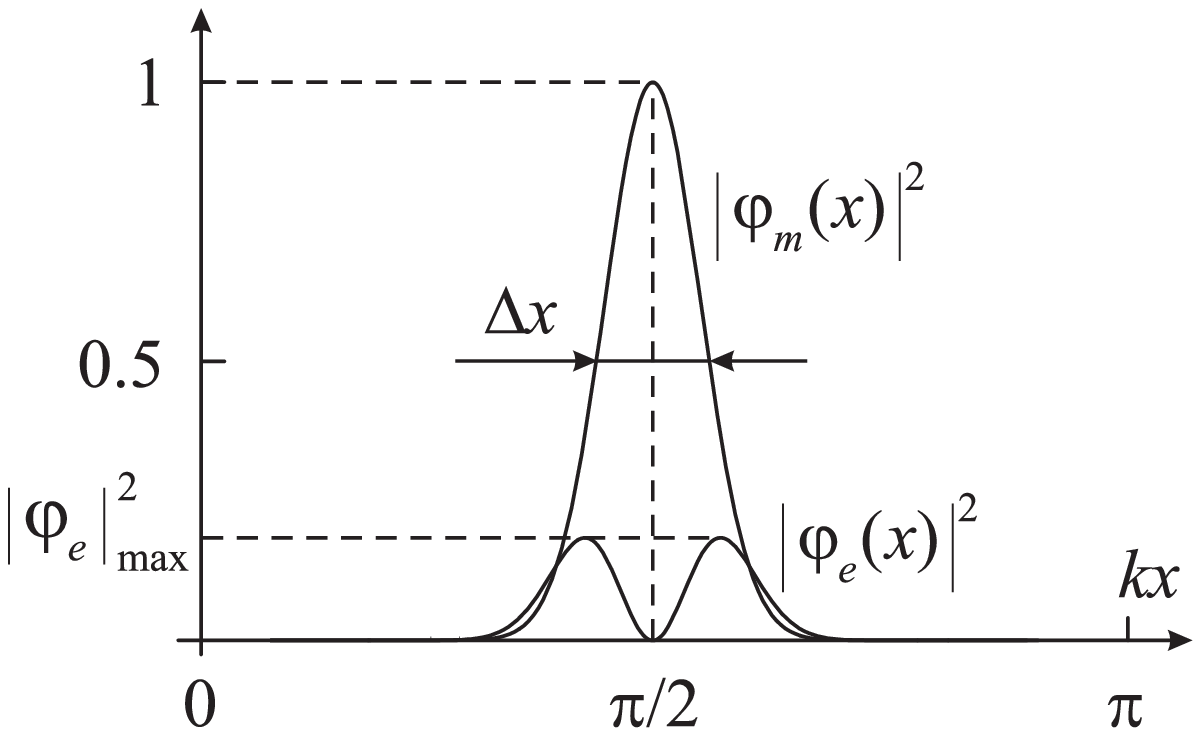}

\noindent  $|\varphi_m(x,t)|^2$ and $|\varphi_e(x,t)|^2$ are not
small is given by
\begin{equation}
 \label{Delta-x}
 \Delta x=\frac{1}{k|\Omega|}\sqrt{\frac{\Gamma}{t}}.
\end{equation}
The condition $\Delta x\ll 1/k$ has the form
\begin{equation}
 \label{small-Delta-x}
 \frac{|\Omega|^2 t}{\Gamma}\gg 1.
\end{equation}
This is the condition under which only relatively narrow regions
close to nodes of a standing light wave give not small
contributions to the total probabilities of scattering $W_{m,
e}(t)$ (\ref{total-prob}).

The parameter $\Delta x$ (\ref{Delta-x}) is one of the key
parameters of the problem under consideration. In particular, the
condition $\Delta x\ll 1/k$ (or the condition given by Eq.
(\ref{small-Delta-x})) justifies the approximation of $\cos^2{kx}$
by a parabolic function in Eqs. (\ref{gamma+}),
(\ref{phi_m_Gauss}), and (\ref{phi_e_Gauss}). As for the used
above approximation $|\Omega|^2\cos^2{(kx)}\ll \Gamma^2/4$ for
$|x-\pi/2k|\lesssim \Delta x$, its validity is related to both
assumptions, $\Delta x\ll 1/k$ and $\Gamma t\gg 1$. Indeed, at
$|x-\pi/2k|\sim\Delta x$, the "effective" position-dependent Rabi
frequency
\begin{equation}
 \label{Rabi-eff}
 \Omega_{eff}(x)\equiv\Omega\cos{(kx)}
\end{equation}
is estimated as
\begin{equation}
 \label{est-Omeg-eff}
 |\Omega_{eff}|\sim|\Omega|k\Delta x\sim\sqrt{\frac{\Gamma}{t}}.
 \end{equation}
The ratio $|\Omega_{eff}|/\Gamma \sim 1/\sqrt{\Gamma t}$ is small
if $\Gamma t\gg 1$ whereas in the opposite case, $\Gamma t\leq 1$,
$|\Omega_{eff}|>\Gamma$ (and, of course, $\Omega_{eff}t\gg 1$).
Hence, we expect that transitions between the metastable and
excited levels have significantly different form at long and short
times, $\Gamma t\gg 1$ and $\Gamma t\leq 1$.  In the first of
these two cases (long-time asymptotic) the transitions
$E_m\rightarrow E_e$ have a form of irreversible transitions to
the quasicontinuum of the wide excited level $E_e$, whereas in the
second case (short-time limit) they take a form of multiple Rabi
oscillations. In the following section we will see how the
dynamics of excitation is affected by a mixture of these two types
of transitions.

The parameter $\Delta x$ (\ref{Delta-x}) determines the relation
between the heights of the curves $|\varphi_m(x, t)|^2$ and
$|\varphi_e(x, t)|^2$: if $|\varphi_m|^2_{max}=1$,
\begin{equation}
 \label{phi_e,max^2}
 |\varphi_e|^2_{max}=
 \frac{|\Omega|^2}{\Gamma^2}k^2\Delta  x^2\, e^{-1}
 =\frac{e^{-1}}{\Gamma t}\ll 1,
\end{equation}
if $\Gamma t\gg 1$.

As $\Delta x$ (\ref{Delta-x}) is the width of the "most important"
region of $x$, where $\Gamma_+$ (\ref{gamma+}) is small and the
corresponding part of atoms decays slowly, we can estimate now a
rigidity of the assumption about the normal incidence of atoms
upon a standing wave. If $p_{0 x}\neq 0$, the atoms move
homogeneously along the $x$-axis. The arising displacement during
the interaction time is $p_{0 x}t/m$, and it must be not larger
than $\Delta x$ to keep atoms decaying slowly, which gives
\begin{equation}
 \label{p_x<?}
 k|v_{0 x}|t\leq \frac{1}{|\Omega|}\sqrt{\frac{\Gamma}{t}}\ll 1,
\end{equation}
where $v_{0 x}=p_{0 x}/m$, and the last inequality follows from
Eq. (\ref{small-Delta-x}).

At last, the definition of the characteristic "important" interval
$\Delta x$ (\ref{Delta-x}) can be used to evaluate the validity
criterion of the adiabatic approximation. In this approximation
the characteristic value of the kinetic energy $-\nabla_x^2/2m$ is
assumed to be small compared to the characteristic value of
$\partial/\partial t$. The latter is estimated as (at $\Gamma t
\gg 1$ and $|x-\pi/2k|\sim\Delta x$): $\partial/\partial
t\sim\Gamma_+\sim|\Omega|^2k^2\Delta x^2/\Gamma\sim 1/t$. In
accordance with the uncertainty principle, we put $\nabla_x\sim
1/\Delta x$, which gives $\nabla_x^2/2m\sim
\omega_r|\Omega|^2t/\Gamma$, where $\omega_r=k^2/2m$ is the recoil
frequency. With the help of these estimates the applicability
criterion of the adiabatic approximation $\nabla_x^2/2m\ll 1/t$
can be reduced to the form
\begin{equation}
 \label{adiabaticity}
 |\Omega|t\ll\sqrt{\Gamma/\omega_r}.
\end{equation}
As, typically, $\omega_r\sim 10^{-3}\Gamma$, this condition can be
fulfilled both at $|\Omega|t<1$ and $|\Omega|t>1$.

By returning to the analysis of the time behavior of the
scattering probabilities, note first that the decay of small
portions of the atomic wave function localized near any given $x$
has an exponential character, though with the decay rate depending
on $x$ (Eqs. (\ref{phi_m_Gauss}) and (\ref{phi_e_Gauss})). But, as
the sum of exponents is not identical to any other exponential
function, it's not surprising that the total probabilities of
scattering (\ref{total-prob}) decay non-exponentially. In other
words, the total probabilities of scattering $W^{(m, e)}_{tot}(t)$
are determined by areas under the curves $|\varphi_{m, e}(x,
t)|^2$ at Fig. 3. With a growing time $t$ these areas shrink, but
the laws of their decreasing are not exponential.

In principle, in the asymptotic limit $\Gamma t\gg 1$ the
integrals in (\ref{total-prob}) are easily calculated with
$\varphi_{m, e}(x, t)$ substituted from Eqs. (\ref{phi_m_Gauss})
and (\ref{phi_e_Gauss}) and the limits of integration extended to
$\mp\infty$. But it's very interesting and instructive to use Eq.
(\ref{gamma+}) before calculations to reduce the integrals over
$x$ to integrals over the narrow-zone width $\Gamma_+$:
\begin{equation}
 \label{w-m-as}
 W_{tot}^{(m)}(t)\approx\frac{\sqrt{\Gamma}}{\pi|\Omega|}
 \int_0^{\infty}\frac{d\Gamma_{+}}{\sqrt{\Gamma_{+}}}\,e^{-\Gamma_{+}t}
 =\frac{\Gamma^{1/2}}{|\Omega|\sqrt{\pi t}}
\end{equation}
and
\begin{equation}
 \label{w-e-as}
 W^{(e)}_{tot}(t)\approx\frac{1}{\pi\Gamma^{1/2}|\Omega|}
 \int\limits_{0}^{\infty}\sqrt{\Gamma_+}\;d\Gamma_+\,e^{-\Gamma_+\,t}
 =\frac{1}{2|\Omega|\sqrt{\pi\Gamma}\, t^{3/2}}.
\end{equation}
So, indeed, the long-time behavior of the total probabilities to
find an atom after scattering at the metastable and excited levels
is determined by power-law rather than exponential dependencies on
the interaction time $t$.

It should be noted, that, in principle, the non-exponential decay
characterized by Eqs. (\ref{w-m-as}), (\ref{w-e-as}) can occur
also in the case of weak Rabi coupling, $|\Omega|<\Gamma$, if only
the conditions (\ref{small-Delta-x}), (\ref{p_x<?}), and
(\ref{adiabaticity}) are fulfilled. The first of these conditions
(\ref{small-Delta-x}) at $|\Omega|<\Gamma$ can be fulfilled only
if the interaction time $t$ is very large,
$t\gg\Gamma/|\Omega|^2\gg 1/|\Omega|\gg1/\Gamma$. Such a long time
can make the restriction of the transverse velocity $v_{0 x}$
(\ref{p_x<?}) too severe to be easily satisfied. For this reason,
the case of strong Rabi coupling looks much more favorable than
the case of weak coupling for observation of the effects described
in this and the following sections. Compatibility of the
conditions (\ref{small-Delta-x}) and (\ref{adiabaticity}) requires
the Rabi frequency to be not too small,
$|\Omega|\gg\sqrt{\omega_r\Gamma}$.

\section{Partial probabilities of scattering into diffraction beams}
To investigate in more details the time evolution of scattering,
let us consider the Fourier transforms of the atomic
center-of-mass wave functions $\varphi_{m,e}(x,t)$
\begin{equation}
 \label{Fourier}
 a_n^{(m,e)}(t)=\frac{k}{\pi}\int_0^{\pi}dx\,\varphi_{m,e}(x,t)\,\exp{(-inx)}.
\end{equation}
The functions $a_n^{(m,e)}(t)$ and their squared absolute values
\begin{equation}
 \label{partial}
 W_n^{(m,e)}(t)=\left|a_n^{(m,e)}(t)\right|^2
\end{equation}
are the probability amplitudes and partial probabilities to find
an atom at the levels $E_m$ or $E_e$ in the $n$-th diffraction
beam with the momentum ${\bf p}_0+n{\bf k}$, which makes an angle
$\theta_n\approx nk/p_0$ with ${\bf p}_0$, $n=0, \pm1, \pm2, ...$.
Equations for $a_n^{(m,e)}(t)$, equivalent to Eqs. (\ref{eq for
Phi}), (\ref{Ham}), are given by
$$
i\dot{a}_n^{(m)}(t)=\left(n^2\omega_r + n\delta \right)
a_n^{(m)}-\frac{\Omega}{4}\left(a_{n-1}^{(e)}+a_{n+1}^{(e)}\right),
$$
\begin{equation}
 \label{gen-system-a_and_b}
 i\dot{a}_{n}^{(e)}(t)=\left(n^2\omega_r + n\delta-\frac{i\Gamma}{2}
 \right)a_n^{(e)}-\frac{\Omega^*}{4}\left(a_{n-1}^{(m)}+a_{n+1}^{(m)}\right).
\end{equation}
These equations look like equations for two coupled anharmonic
oscillators in a resonance field with the resonance detuning
$\delta=kp_{0x}/m$ and the anharmonicity parameter coinciding with
the recoil frequency $\omega_r$. The above-discussed adiabatic
approximation corresponds to ignoring the anharmonicity terms
$n^2\omega_r$ in Eqs. (\ref{gen-system-a_and_b}). With the
additional assumption about the normal incidence, $p_{0x}=0$, Eqs.
(\ref{gen-system-a_and_b}) take the form
$$
i\dot{a}_n^{(m)}(t)=-\frac{\Omega}{4}\left(a_{n-1}^{(e)}+a_{n+1}^{(e)}\right),
$$
\begin{equation}
 \label{system-a and b adiab}
 i\dot{a}_{n}^{(e)}(t)=-\frac{i\Gamma}{2}a_n^{(e)}
 -\frac{\Omega^*}{4}\left(a_{n-1}^{(m)}+a_{n+1}^{(m)}\right).
\end{equation}
The transition from Eqs. (\ref{gen-system-a_and_b}) to
(\ref{system-a and b adiab}) can be considered as the Raman-Nath
approximation for the two-dimensional (2D) system. It should be
noted, however that both Eqs. (\ref{system-a and b adiab}) and
their solutions presented below differ significantly from and are
much more complicated than the standard 1D Raman-Nath equation and
its solution \cite{BW}.

The simplest way of finding $a_n^{(m,e)}(t)$ obeying Eqs.
(\ref{system-a and b adiab}) is related to the calculation of the
Fourier transforms (\ref{Fourier}) of the earlier found functions
$\varphi_m(x, t)$ (\ref{fi_m}) and $\varphi_e(x, t)$ (\ref{fi_m}).
Not dwelling upon the details of calculations, let us present here
the arising results:
$$
 W_{2n}^{(m)}(t)=e^{-\frac{\Gamma t}{2}}
 \left|J_{2n}\left(\frac{|\Omega|t}{2}\right)+\frac{\Gamma
 t}{4}\int_0^1 dz J_{2n}\left(\frac{|\Omega|t}{2}z\right)\right.
$$
\begin{equation}
 \label{w^m_2n(t)}
 \times\left.
 \left[\frac{I_1\left(\frac{\Gamma t}{4}\sqrt{1-z^2}\right)}{\sqrt{1-z^2}}
 +I_0\left(\frac{\Gamma t}{4}\sqrt{1-z^2}\right)\right]\right|^2
\end{equation}
and
$$
W_{2n+1}^{(e)}=\left(\frac{|\Omega|t}{4}\right)^2
\,\exp\left(-\frac{\Gamma t}{2}\right) \Biggl|\int_{0}^{1} dz\;
I_0 \left(\frac{\Gamma
 t}{4}\sqrt{1-z^2}\right)
$$
\begin{equation}
 \label{w^e_2n+1(t)}
 \times\left[J_{2n+2}\left(\frac{|\Omega|t}{2}z\right)
 -J_{2n}\left(\frac{|\Omega|t}{2}z\right)\right]\Biggr|^2,
\end{equation}
whereas $W_{2n+1}^{(m)}(t)\equiv 0$ and $W_{2n}^{(e)}(t)\equiv 0$.
In Eqs. (\ref{w^m_2n(t)}), (\ref{w^e_2n+1(t)}) $J_{2n}$ are the
Bessel functions and $I_0$ and $I_1$ are the modified Bessel
functions \cite{AS}. It should be emphasized that in derivation of
Eqs. (\ref{w^m_2n(t)}) and (\ref{w^e_2n+1(t)}) we did not make any
assumptions about a value of the parameter $\Gamma t$ and, hence,
these equations are valid both at short interaction times, $\Gamma
t\leq 1$, and in the long-time asymptotic limit, $\Gamma t\gg 1$.
The functions $W_{2n}^{(m)}(t)$ (\ref{w^m_2n(t)}) and
$W_{2n+1}^{(e)}(t)$ (\ref{w^e_2n+1(t)}) are plotted in Fig. 4a and
b. In contrast to the

\vskip 12pt

\inputEps{220pt}{\,\,}{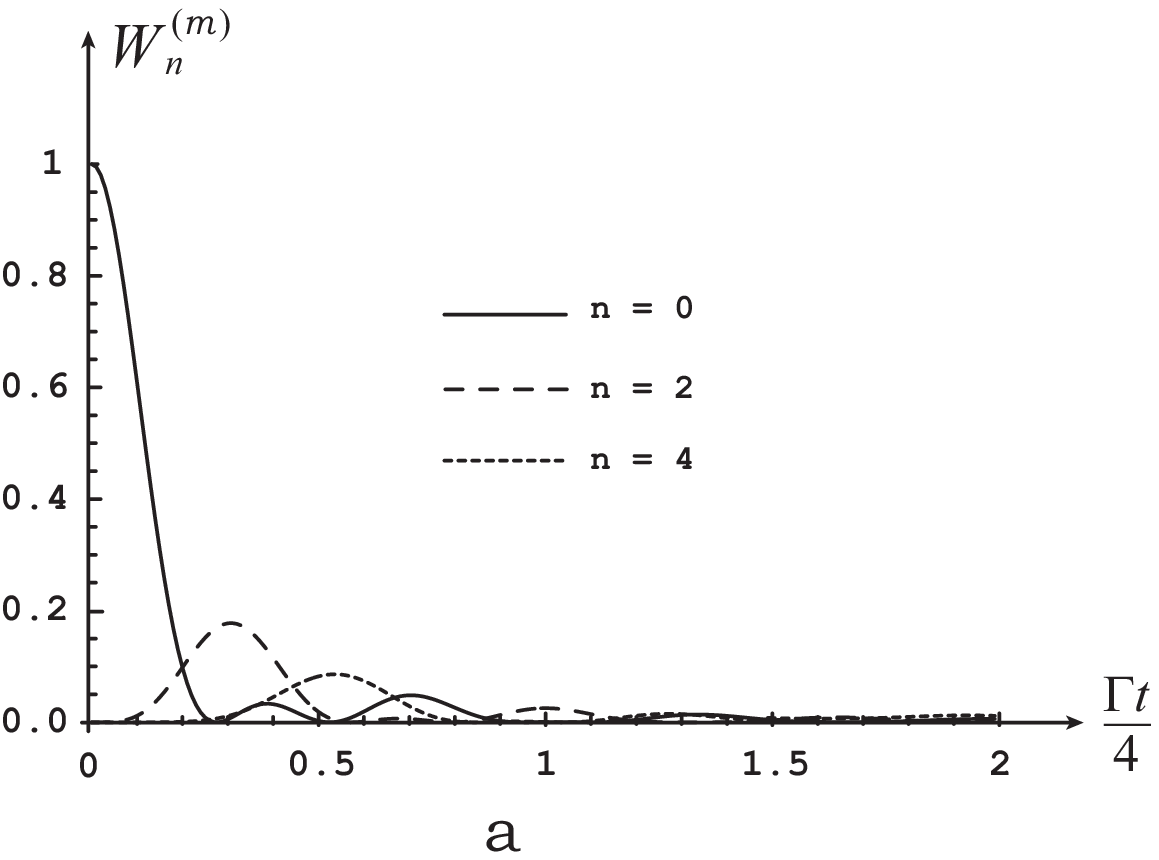}

\vskip 12pt

\inputEps{220pt}{Figure 4. The time-dependent partial probabilities
$W_n^{(m)}(t)$ (a) and $W_n^{(e)}(t)$ (b);
$|\Omega|/\Gamma=5$.}{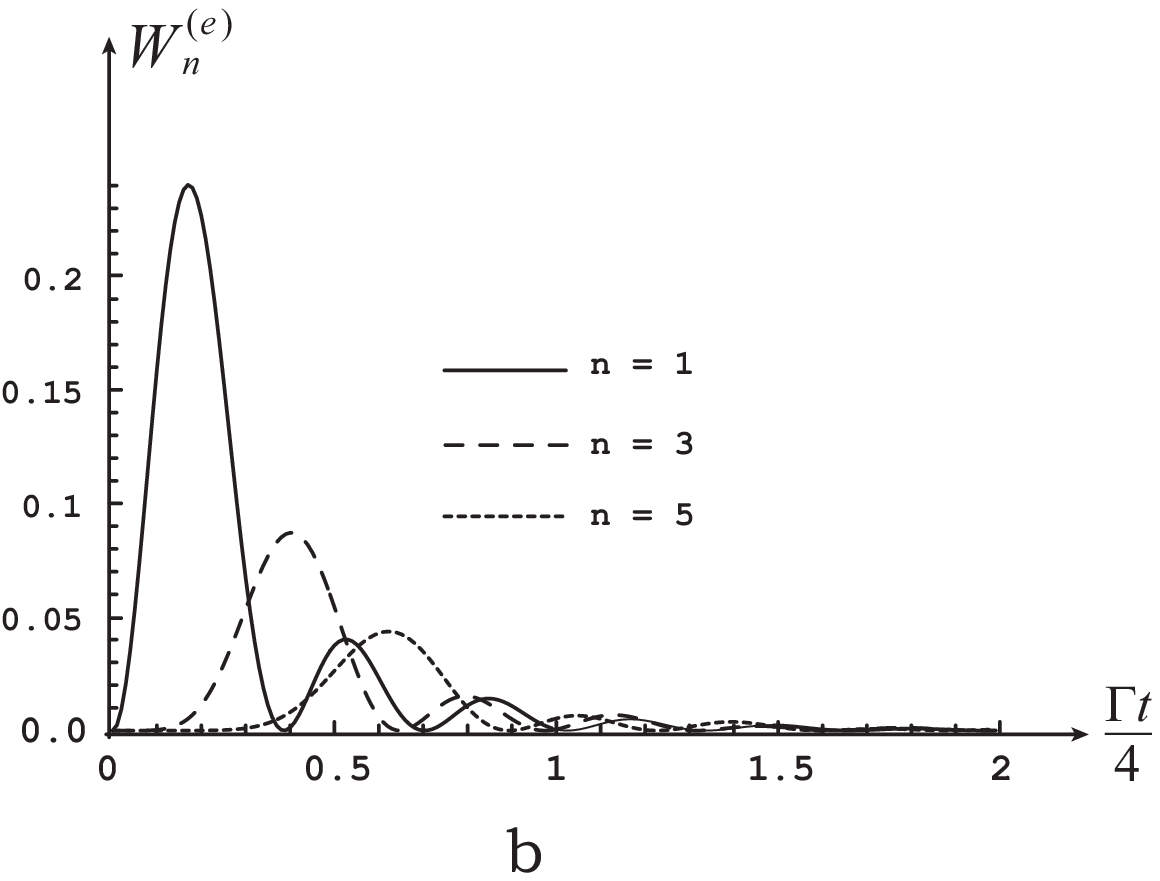}

\noindent well-known Rabi oscillations in a pure two-level system
in a resonance field, oscillations of $W_{2n}^{(m)}(t)$ are
aperiodic and both positions of peaks and zeros of
$W_{2n}^{(m,e)}(t)$ depend on $n$.

Alternatively to (\ref{total-prob}), the total probabilities of
finding atoms after scattering at the levels $E_m$ and $E_e$ can
be defined as sums over the diffraction beams
\begin{equation}
 \label{total-prob-sum}
 W_{tot}^{(m,e)}(t)=\sum_n\mid a_n^{(m,e)}(t)\mid^2.
\end{equation}

The dependencies $W_{tot}^{(m)}(t)$ and $W_{tot}^{(e)}(t)$
(\ref{total-prob-sum}) calculated numerically by summing
$W_{2n}^{(m)}(t)$  (\ref{w^m_2n(t)}) and $W_{2n+1}^{(e)}(t)$
(\ref{w^e_2n+1(t)}) are shown in Figs. 5a and b. In contrast to
$W_{2n}^{(m)}(t)$ and $W_{2n+1}^{(e)}(t)$, oscillations of
$W_{tot}^{(m)}(t)$ and $W_{tot}^{(e)}(t)$ are periodic, and their
period coincides with that of the Rabi oscillations in a pure
two-level system driven by a resonant field with the
field-strength amplitude $2{\bf E}_{0}$ (the dashed curves in Fig.
5) . However, as it's seen well from Fig. 5, the quasi-Rabi
oscillations of the functions $W_{tot}^{(m)}(t)$ and
$W_{tot}^{(e)}(t)$ for scattered atoms are strongly suppressed
compared to those of a two-level system. The effect of suppression
of Rabi oscillation is explained mainly by a kind of inhomogeneous
broadening. Oscillations of partial probabilities
$W_{2n}^{(m)}(t)$ (\ref{w^m_2n(t)}) and $W_{2n+1}^{(e)}(t)$
(\ref{w^e_2n+1(t)}) are well pronounced and their amplitudes are
large enough (Fig. 4). However, as "periods" of these oscillations
are different for different $n$, summation over $n$ smoothes over
the oscillations and decreases their amplitudes. In insets of

\vskip 12pt

\inputEps{220pt}{\,\,}{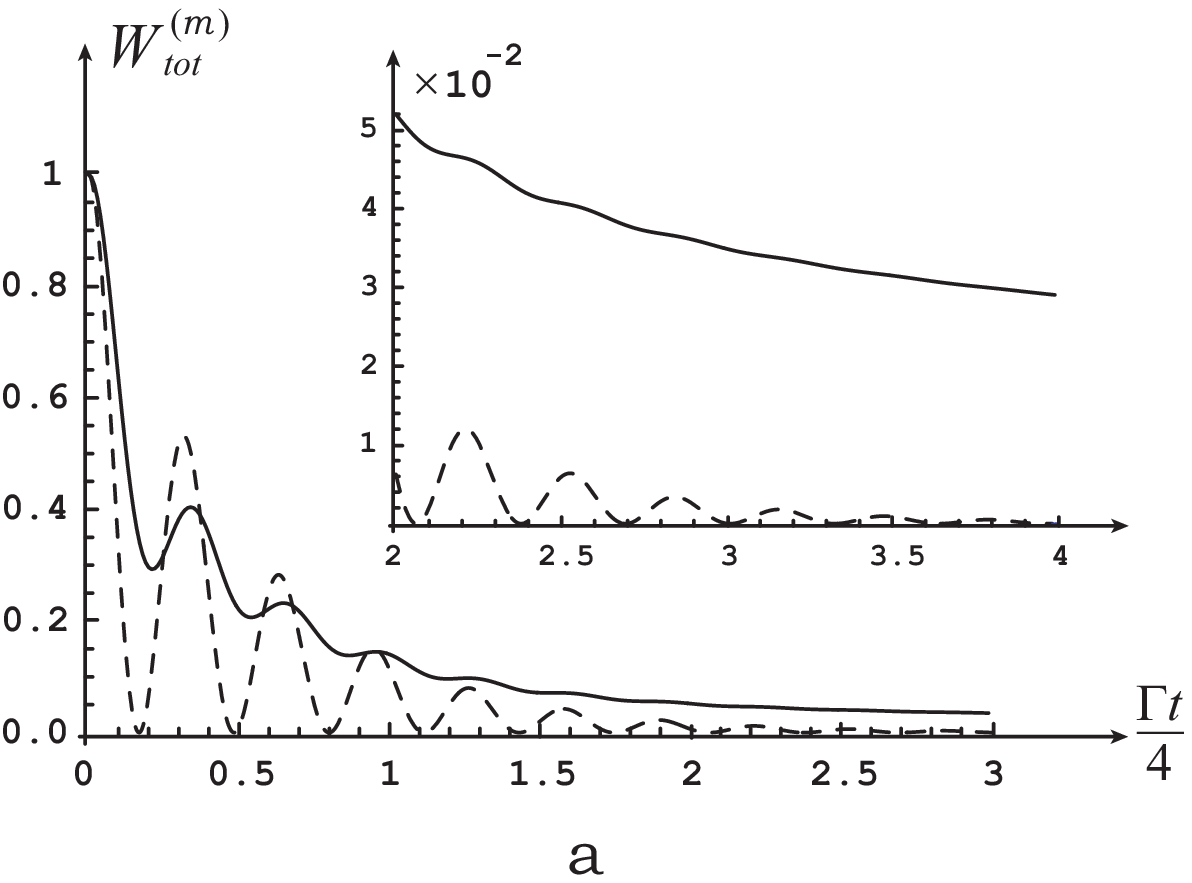}

\inputEps{220pt}{Figure 5. Total populations $W_{tot}^{(m)}(t)$
(a) and $W_{tot}^{(e)}(t)$ (b) at $|\Omega|/\Gamma=5$ (solid
lines) and the pure two-level-system population (dashed
curve)}{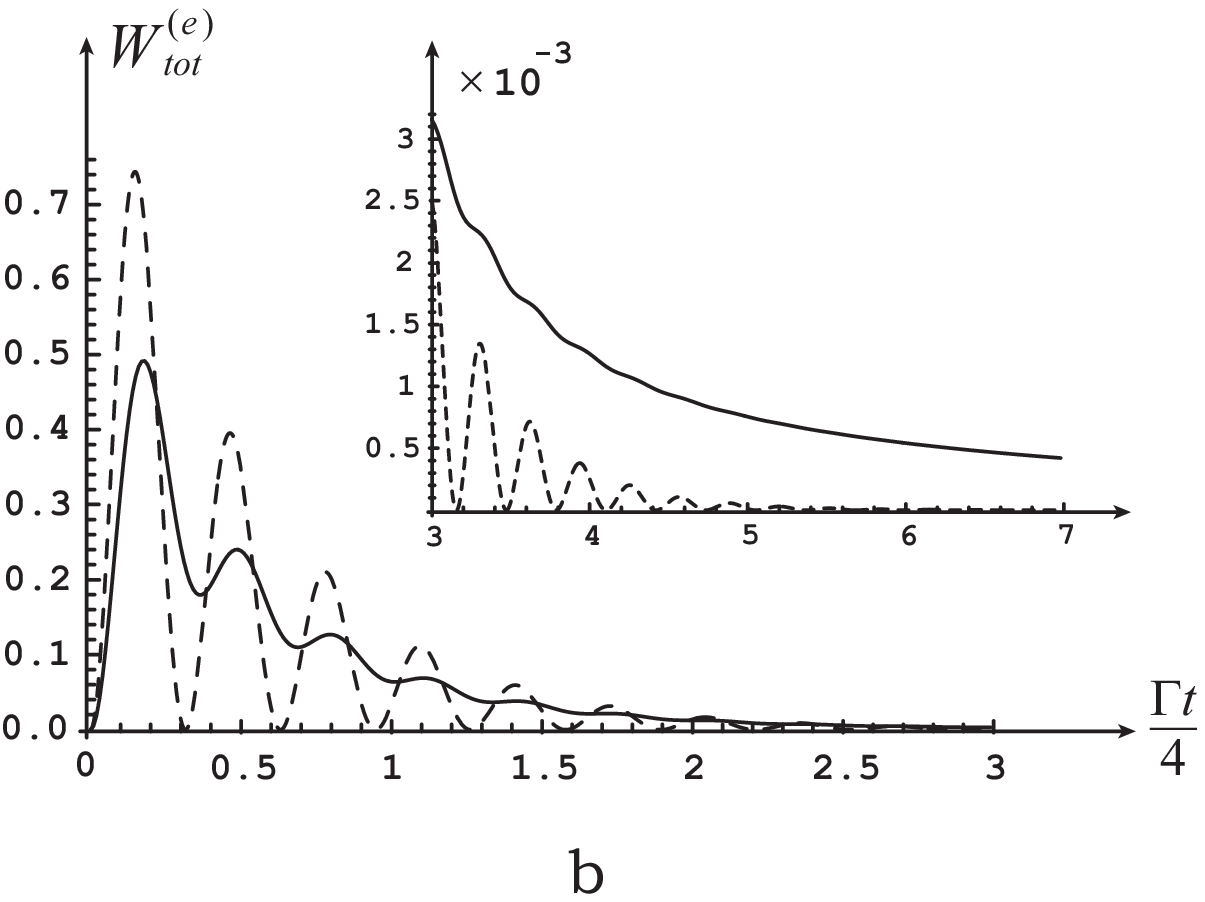}

\noindent Fig. 5 it's seen clearly that the degree of suppression
the Rabi oscillations increases with a growing value of $\Gamma
t$: in the case $\Gamma t \gg 1$ oscillations are almost
completely smoothed away. This effect is related to the discussed
above change in the character of transitions between the levels
$E_m$ and $E_e$: Rabi oscillations in the case $\Gamma t\lesssim
1$ and "discrete level - quasicontinuum" transitions in the case
$\Gamma t\gg 1$.

Another effect which is clearly seen in the insets of Figs. 5a and
b is the scattering-induced slowing down of the radiative decay.
The long-time behavior of the probabilities $W_{tot}^{(m,e)}(t)$
is described rather well by asymptotic formulas of Eqs.
(\ref{w-m-as}) and (\ref{w-e-as}). The difference with exponential
decay in a pure two-level system is rather well pronounced.

\section{Conclusion}
So, the main predictions of the carried out consideration are (i)
suppression of the Rabi oscillations in the case of atom
scattering (compared to a pure two-level system) and (ii) slowing
down the radiative decay and formation of nonexponential
(power-law) tails in the dependencies $W_{tot}^{(m,e)}(t)$.
Qualitatively, our interpretation of these effects consists of the
following. The Rabi oscillations are suppressed owing to
inhomogeneous-broadening-like effects when the partial
probabilities of scattering are summed over the diffractions
beams. The power-law dependencies and slowing down of the
radiative decay arise because of the position-dependent modulation
of the field-strength amplitude in a standing light wave and,
hence, the position-dependent modulation of the decay rate of
slowly decaying quasienergy atomic levels $\Gamma_+(x)$. As the
result, of this modulation the manifold of the arising quasienergy
levels is characterized by continuously varying with $x$ and
approaching zero at $x=\pi/2k$ width. Populations at these
quasienergy levels decrease exponentially but with different,
$x$-dependent rates, and their superposition gives rise to the
non-exponential decay laws.

We assume that observation of these effects can be made in the
framework of an experiment similar to \cite{Z, Z2} though with
some modifications, to provide the conditions for the strong Rabi
coupling and the normal-incidence diffraction regime of
scattering. To compare directly the time evolution of atomic
populations in atoms scattered by a standing light wave and in a
pure two-level system, one can make two series of similar
measurements: in a standing light wave and in a single travelling
wave of a doubled field strength amplitude. In the last case the
standing-wave scattering effects disappear and the large-amplitude
Rabi oscillations and the usual exponential decay have to be
observed.

Finally, the above-described scattering-induced suppression of the
radiative decay in the case of strong Rabi coupling reminds the
effect of interference stabilization in Rydberg atoms in a strong
light field \cite{Movs}, \cite{book}. In both cases the effects of
slowing down the decay processes are related to formation of
narrow quasienergy levels and interference of transitions from
different levels to the common continuum. We find this analogy
important because it establishes links between different regions
of physical phenomena and demonstrates a rather general character
and fruitfulness of the idea of interference stimulated by
sufficiently strong interactions.

\section{Acknowledgement}
The work is supported partially by the Russian Foundation for
Basic Research (grant $\#$ 02-02-16400). M. Fedorov gratefully
acknowledges the support of the Humboldt foundation and the great
hospitality enjoyed at the Abteilung f\"ur Quantenphysik,
University of Ulm, Germany. The work of W. P. Schleich is
partially supported by DFG.

\newpage

\end{document}